\documentclass[final,5p,twocolumn]{elsarticle}

\usepackage{amssymb}
\usepackage{amsmath}
\usepackage{epsfig}

\biboptions{sort}

\journal{Chemical Physics Letters}

\begin{document}

\begin{frontmatter}



\title{State-selected ion-molecule reactions \\ with Coulomb-crystallized molecular ions in traps}


\author{Xin Tong}
\author{Tibor Nagy}
\author{Juvenal Yosa Reyes}
\author{Matthias Germann}
\author{Markus Meuwly\corref{cor}}
\author{Stefan Willitsch\corref{cor}}
\cortext[cor]{Corresponding authors. E-mail: m.meuwly@unibas.ch, stefan.willitsch@unibas.ch}

\address{Department of Chemistry, University of Basel, Klingelbergstrasse 80, 4056 Basel, Switzerland}

\begin{abstract}
State-selected Coulomb-crystallized molecular ions were employed for
the first time in ion-molecule reaction studies using the prototypical charge-transfer process $\mathrm{N_2^++N_2\rightarrow N_2+N_2^+}$ as an example. By preparing the reactant ions in a
well-defined rovibrational state and localizing them in space
by sympathetic cooling to millikelvin temperatures in an ion trap,
state- and energy-controlled reaction experiments with sensitivities
on the level of single ions were performed. The
experimental results were interpreted with quasi-classical trajectory simulations on a six-dimensional
potential-energy surface which provided detailed insight into translation-to-rotation energy transfer
occurring during charge transfer between N$_2$ and N$_2^+$. 
\end{abstract}


\end{frontmatter}


\section{Introduction}
\label{intro}

Achieving full control over the internal quantum states as well as the
collision energy of the reaction partners to probe the detailed
state-to-state dynamics and energy dependence of chemical processes has been
one of the long-standing aspirations in gas-phase chemical studies. In
the past, a range of experimental methods has been developed to
address these objectives, e.g., internal cooling of the reactants in
supersonic expansions, experiments with crossed molecular beams at
variable relative velocities and laser preparation of the reactants in specific
quantum states (see, e.g., Refs. \cite{liu06a, teslja06a, yang11a} and
literature cited therein). Very recently, new techniques have been
established which allow the preparation of molecules at extremely low
translational temperatures $< 1$~K and at the same time enable an
unprecedented degree of control over their kinetic energy
\cite{carr09a, schnell09a, willitsch12a}. In conjunction with the
simultaneous preparation of the molecules in well-defined
rotational-vibrational and even hyperfine states, ''cold molecules''
methods are now starting to pave the way for studies of collisional
processes and chemical reactions in new physical regimes and at levels
of detail and sensitivity which have not been possible before
\cite{gilijamse06a, willitsch08a, ospelkaus10b, demiranda11a,
  hall11a}.

Whereas most efforts have so far concentrated on neutral molecules,
techniques for cooling and controlling molecular ions have recently
made impressive progress as well. From an experimental perspective,
the sensitivity of the ion motion to weak stray electric fields in the
apparatus renders the precise control of their kinetic energy
difficult. This problem can be overcome by trapping the ions and
cooling them sympathetically to millikelvin temperatures by the
interaction with co-trapped laser-cooled atomic ions
\cite{molhave00a}. Under these conditions, the ions localize in space
to form ordered structures usually referred to as "Coulomb crystals"
in which it is possible to observe, manipulate and address single ions
\cite{willitsch08b, willitsch12a}.

Moreover, the preparation of molecular ions in well-defined quantum
states is experimentally challenging and indeed only a handful of
reaction studies with rotationally state-selected ions have been
reported thus far (see, e.g., \cite{mackenzie94a, glenewinkel97a,
  green00a, dressler06a, paetow10a}). Only very recently it has become
possible to prepare Coulomb-crystallized molecular ions in
well-defined internal quantum states \cite{tong10a, tong11a,
  staanum10a, schneider10a}, so that a simultaneous control over the
kinetic energies, positions and internal states of the ions can now be achieved.

In the present article, we report the first study of a chemical
reaction with rovibrationally state-selected Coulomb-crystallized
molecular ions. We investigate the prototypical symmetric
charge-transfer (CT) reaction $\mathrm{N_2^++N_2\rightarrow
  N_2+N_2^+}$ between state-prepared sympathetically-cooled N$_2^+$
ions and internally cold N$_2$ molecules from a supersonic expansion
at a well-defined collision energy.

The mechanism and kinetics of this reaction have been studied
previously, see, e.g., Refs. \cite{friedrich84a, vankoppen84a,
  frost94a, kato98a, sohlberg99a, frost01a} and references therein. At
low energies, the reaction was found to proceed via a long-lived
N$_4^+$ reaction complex which forms at the collision (Langevin) rate
\cite{vankoppen84a, frost94a, frost01a}. Using isotopically labeled
nitrogen ions, Frost et al. showed that near-thermal collisions
between N$_2^+$ ions and N$_2$ molecules in their vibrational ground
states lead to a symmetric CT with a rate constant of
$k=4.2\times10^{-10}$~cm$^3$~s$^{-1}$ which amounts to one half of the
collision rate constant \cite{frost94a}. This observation was
rationalized in terms of a symmetric sharing of the charge between the
two N$_2$ moieties in the N$_4^+$ reaction complex so that the
probability for CT amounts to 50\% upon its
breakup. Vibration-to-vibration (V-V) and vibration-to-translation
(V-T) energy transfer occurring during CT have been
investigated in this system both experimentally and theoretically, see
Refs. \cite{mcafee81a, sohlberg91a, frost94a, kato98a, sohlberg99a}
and references therein. Kato et al. observed multi-quantum
vibrational-energy transfer at near-thermal collision energies which
was rationalized in terms of energy redistribution within the strongly
bound N$_4^+$ reaction complex \cite{kato98a}. Also, evidence
for vibration-to-rotation and translation (V-R,T) energy transfer was
found.

To our knowledge, the present work is the first to achieve rotational
state selection of the reactant ions and rotational resolution in the
analysis of the product-ion state populations allowing to gain insight
into the dynamics of translation-to-rotation (T-R) energy transfer
occurring during CT. The Coulomb-crystal technology
employed in the present study enabled us to monitor reactive
collisions in small ensembles of 20 to 30 state-selected, spatially
localized N$_2^+$ ions with single-particle sensitivity. The
experimental results were interpreted with quasi-classical trajectory
calculations on a fully-dimensional potential-energy surface (PES)
which provided insight into the charge- and energy-transfer dynamics
underlying this fundamental reaction.


\section{Experimental methods}
\label{expt}

\begin{figure}[t]
\begin{center}
\epsfig{file=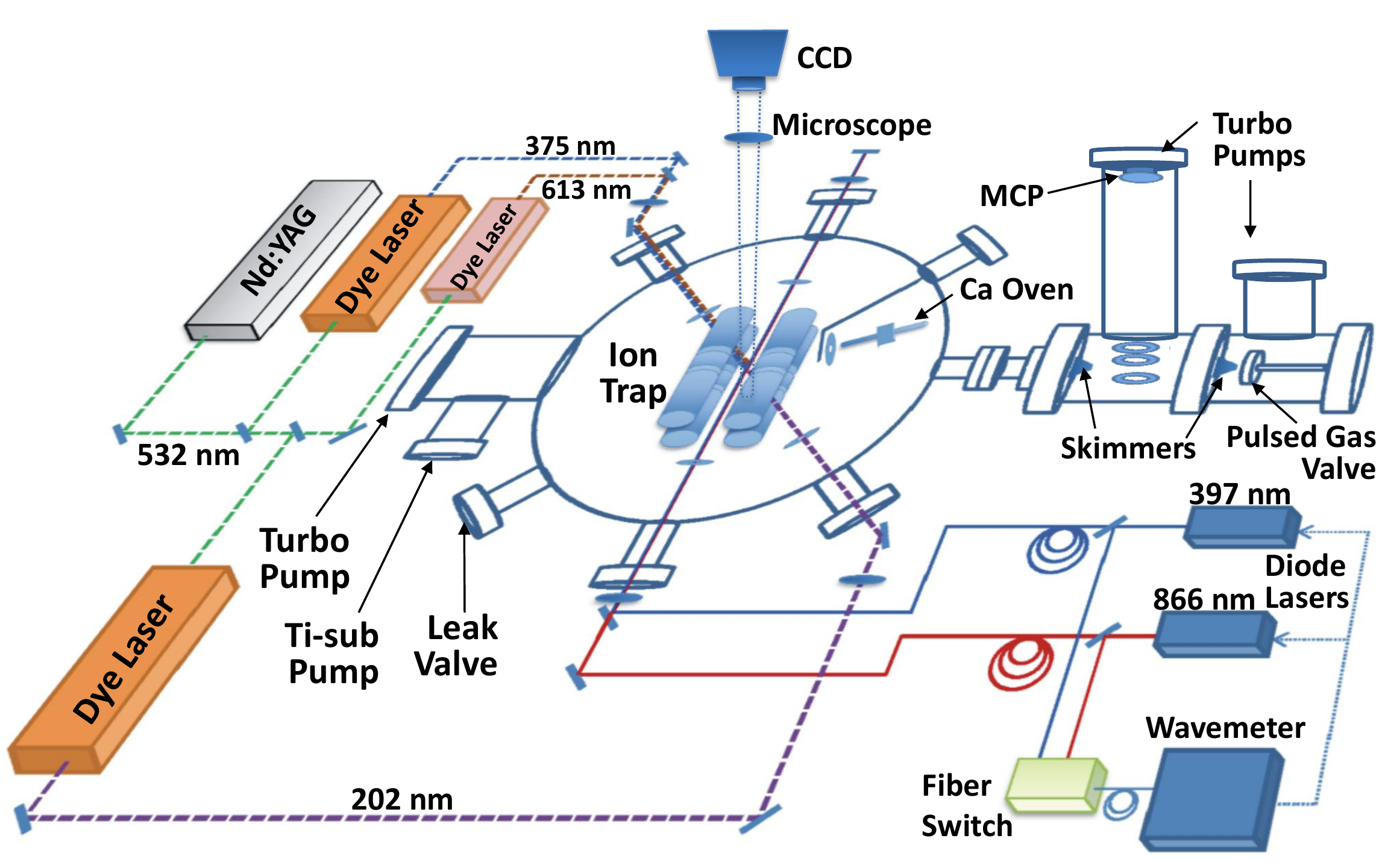,width=0.9\columnwidth}
\end{center}
\caption{Schematic of the experimental setup. }
\label{expt_fig}
\end{figure}

\begin{figure}[t]
\begin{center}
\epsfig{file=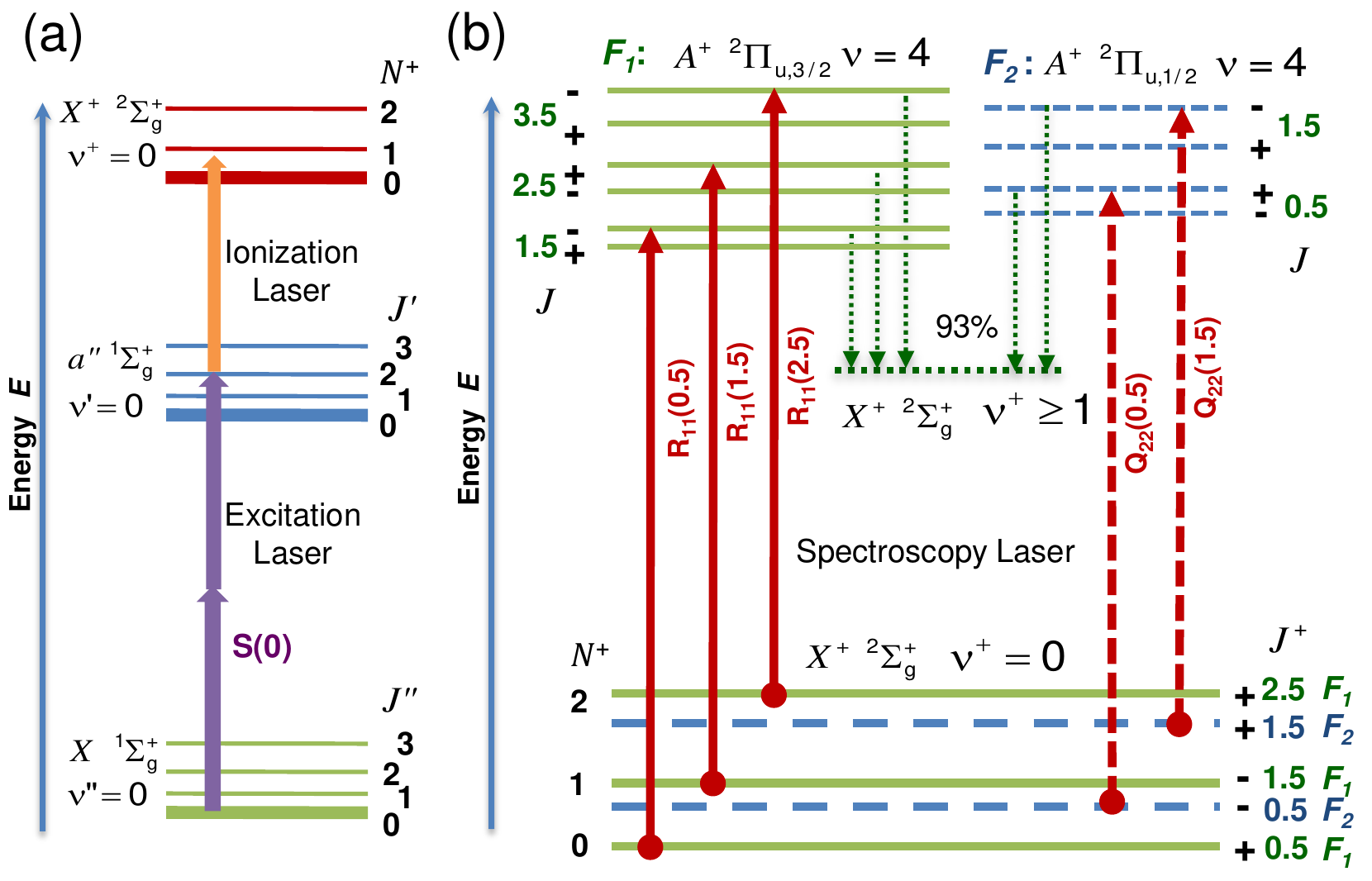,width=0.9\columnwidth}
\end{center}
\caption{(a) Resonance-enhanced threshold-photoionization scheme used
  to generate N$_2^+$ ions in the $X^+~^2\Sigma_g^+,v^+=0, N^+=0, F_1$
  spin-rovibrational ground state. (b) Laser-induced charge-transfer
  (LICT) scheme for probing the populations in the spin-rotational
  levels $N^+, F_{1,2}$ of the vibronic ground state of N$_2^+$ produced by CT reactions. See
  text for details.}
\label{pi}
\end{figure}

The experimental setup for generating quantum-state selected
Coulomb-crystallized N$_2^+$ ions (see Fig. \ref{expt_fig}) has been
described in detail in previous publications \cite{tong10a,
  tong11a}. Briefly, N$_2^+$ ions in the rovibronic ground state
$X^+~^2\Sigma_g^+,v^+=0, N^+=0, F_1$ were prepared by a $[2+1^\prime]$
resonance-enhanced threshold-photoionization sequence via the
$a^{\prime\prime}~^1\Sigma_g^+,v^\prime=0,J^\prime=2$ intermediate
level of neutral N$_2$ as shown in Fig. \ref{pi} (a). Here, $v^+$ and
$v^\prime$ ($N^+$ and $J^\prime)$ stand for the vibrational
(rotational) quantum numbers of the cationic ground and neutral
intermediate states, respectively, and $F_1$ denotes the
spin-rotational component. The N$_2^+$ ions were generated inside a
linear quadrupole ion trap \cite{willitsch08b, willitsch12a} from a
collimated, pulsed supersonic molecular beam of pure N$_2$ gas. The
molecular beam was positioned at a distance of $\approx400~\mu$m from
the Coulomb crystal in order to prevent collisions between
ions and neutrals during the loading phase. The rotational temperature
of the N$_2$ molecules in the beam was determined to be $T_\text{rot}
\approx 10$~K by resonance-enhanced multiphoton-ionization (REMPI)
spectroscopy \cite{tong11a} corresponding to populations of
50, 25, 22, and 3~\% in the $J=0,1,2,3$ rotational states. The beam
velocity was estimated to be $\approx 787$~ms$^{-1}$ from
flow-dynamics models \cite{miller88a}.

Immediately after their generation, typically 25 state-selected
N$_2^+$ ions were sympathetically cooled to translational
temperatures of $\approx 10$~mK by the interaction with laser-cooled
Ca$^+$ ions to form bi-component Coulomb crystals \cite{willitsch08b,
  willitsch12a}. Ca$^+$ ions were produced in the center of the trap
by non-resonant photoionization of Ca atoms emanating from a Ca
oven. The Ca$^+$ ions were laser cooled on the
$4s~^2S_{1/2}\rightarrow4p~^2P_{1/2}$ transition using diode-laser
radiation at 397~nm. Another diode-laser beam at 866~nm was used to
repump population on the $3d~^2D_{3/2}\rightarrow4p~^2P_{1/2}$
transition to close the laser cooling cycle. The resulting
Ca$^+$/N$_2^+$ bi-component crystals were imaged by collecting the
spatially resolved laser-cooling fluorescence of the Ca$^+$ ions using
a microscope coupled to camera. In the images, the positions of the
non-fluorescing molecular ions are visible as a dark region in the
center of the crystals, see Fig. \ref{LICT} (a).

Reactive collisions between state-selected N$_2^+$ ions and neutral N$_2$
molecules were initiated by overlapping the molecular beam with the
bi-component Coulomb crystal immediately after ion loading and
sympathetic cooling. The collision energy
$E_\text{col}\approx$0.045~eV in the present experiments was entirely
dominated by the kinetic energy of the N$_2$ molecules in the beam. 
Neglecting the small initial rotational excitation of neutral N$_2$,
the maximum kinetic energy of the product ions cannot exceed the
total kinetic energy of the reactants which is much smaller than the trap
depth ($>$2~eV). Consequently, the product ions remained trapped and
were sympathetically re-cooled into the Coulomb crystal. Because the
product ions were chemically identical to the reactant ions, their
presence manifested itself in a time-dependent increase of the
population in rotationally excited states in the ensemble of
sympathetically-cooled N$_2^+$ ions. Vibrational excitation of the
products was precluded on energetic grounds. Rate constants were determined by measuring the N$_2^+$ spin-rotational state populations as a function of the effective time of reaction with N$_2$ molecules from the beam and fitting the results to a kinetic model as detailed in
Sec. \ref{results}.

The rotational-state populations of the N$_2^+$ product ions were
probed by optical pumping to vibrationally excited levels to promote
CT reactions with Ar atoms which are energetically forbidden in the
vibrational ground state (laser-induced charge-transfer (LICT)
spectroscopy \cite{tong10a, tong11a}). LICT was initiated by exciting
transitions to selected rotational levels of the
$X^+~^2\Sigma^+_g,v^+=0\rightarrow A^+~^2\Pi_u, v=4$ state with a
pulsed dye laser operating at 613~nm, see Fig. \ref{pi}
(b). Subsequent fluorescent decay lead to the population of
vibrationally excited states $v^+\geq 1$ in the electronic ground
state with a probability of 93\%. The removal of vibrationally excited
N$_2^+$ ions by CT with Ar gas leaked into the chamber was directly
observed by the reduction of the core of non-fluorescing ions in the
images, see Fig. \ref{LICT} (b). The number of N$_2^+$ ions lost by
LICT represented a direct measure of the population in the initially
excited spin-rotational state. This number was determined by a
comparison of the experimental fluorescence images with simulated
images generated from molecular-dynamics simulations \cite{tong10a,
  tong11a}, see Fig. \ref{LICT} (a).


\section{Theoretical methods}

\subsection{Potential energy surfaces}
\label{theo_pes}

\begin{figure}[t]
\begin{center}   
\epsfig{file=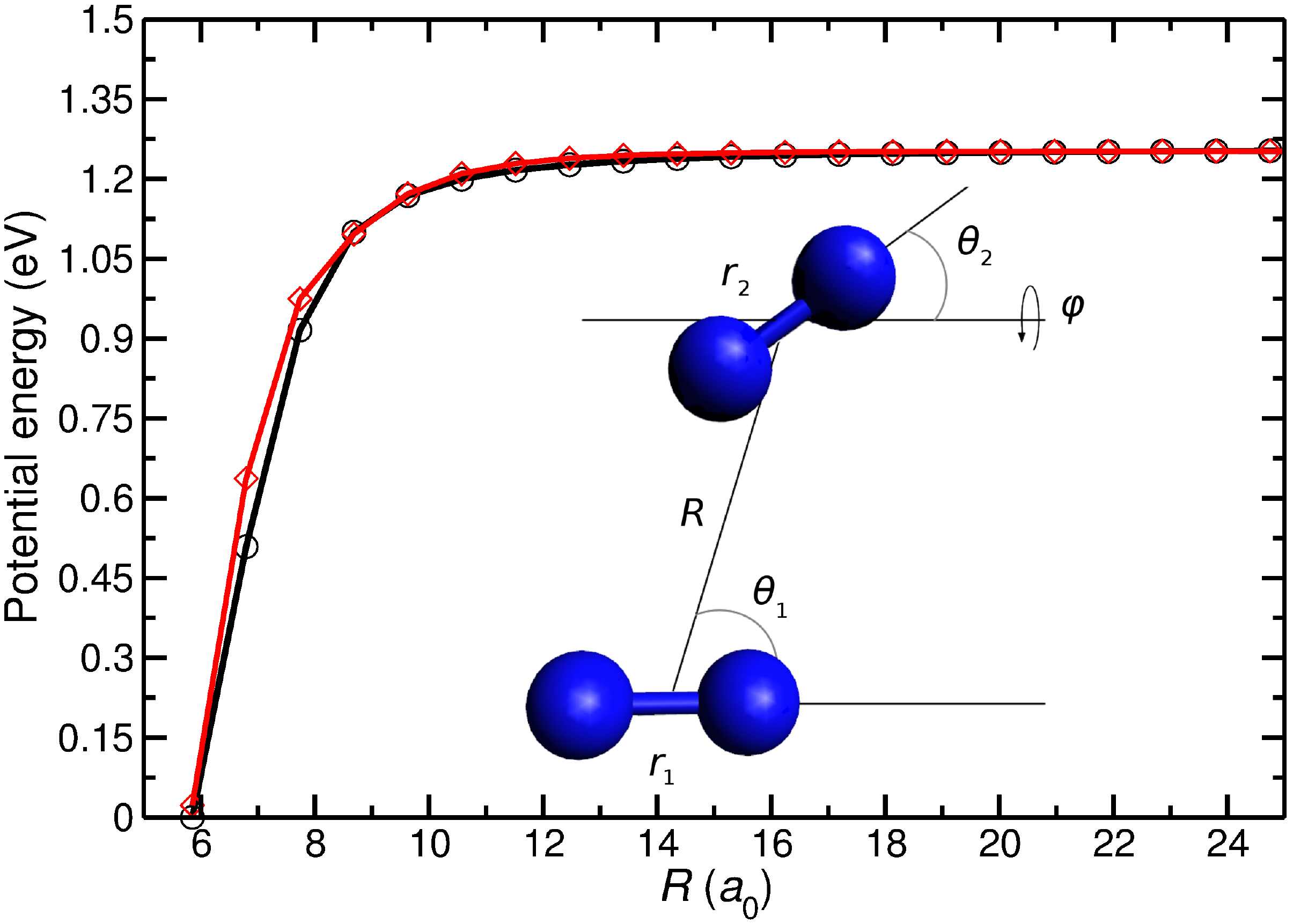,width=\columnwidth}
\end{center}                                
\caption{Comparison of the UCCSD/cc-pVTZ (black) and fitted energies (red) for linear structures with $r_1=r_2=2.08~a_0$ as a function of $R$. The inset shows
  the relevant coordinates scanned in the ab initio
  calculations.}
  \label{irc}
\end{figure}

The ground state six-dimensional potential-energy surface (PES) of the
N$_2$-N$_2^+$ system was computed at the UCCSD level \cite{Purvis:1982,
  Scheiner:1987} with Dunning's correlation consistent polarized
valence basis set (cc-pVTZ) \cite{Dunning:1989} using
Gaussian03 \cite{Gaussian03:1}. Single-point calculations were performed
for 5565 non-equivalent geometries with energies up to 1.9~eV relative to the minimum of the N$_4^+$ complex on a non-equidistant rectangular
grid including the N-N separations $r_1$ and $r_2$ in N$_2$ and
N$_2^+$, respectively, the center of mass distance $R$ between the two
diatomic molecules, and the angles $\theta_1$, $\theta_2$ and $\phi$
(see the inset in Fig. \ref{irc}). The
calculated well-depth ($D_{\rm e}$) of the complex is 1.254~eV, in
good agreement with previous even higher-level theoretical estimates (1.26~eV at the
(RCCSD(T)/vqz(spdfg) level of theory \cite{leonard:1999}). For comparison, the experimentally determined dissociation energy of N$_4^+$ is $D_0=1.06$~eV \cite{weitzel02a}.

The global PES was represented by three surfaces, referred to as one
``bound'' (N$_4^+$) and two ``unbound'' ones (N$_2$--N$_2^+$ and
N$_2^+$--N$_2$). They were distinguished by the distance $R$ and the
localisation of the majority of the charge.  For $R \leq 7.09~ a_0~ (3.75$~\AA)
the system is considered to be bound whereas for $R \geq 7.56~a_0~ (4.0$~\AA) it
is unbound. The intermediate region ($7.09-7.56~a_0$) represents the
transition region during the dynamics, where the surfaces are
connected by a smooth
switching-function \cite{Johnson:1985}.

To allow a) bond formation between any two atoms of N$_2$ and N$_2^+$ and
b) dissociation of the complex into either the charge-preserving or
charge-transferring state at least 8 force fields (FFs) are needed for the bound
state and 4 FFs are necessary for each of the unbound states. These
FFs are related to each other through permutation of their
parameters. Within each set of FFs always the lowest energy
surface is followed, except for when they are close in energy (within
$\approx 0.02$~eV) in which case they are smoothly joined by an energy
difference-based switching function \cite{Johnson:1985}.

Morse and Lennard-Jones potentials were used for the bond stretches
and the intermolecular interactions, respectively, for the unbound
states. Morse parameters for the isolated molecules were determined
from PES scans at the UCCSD level of theory. For the bound state,
Morse potentials were used for all bonds of the complex and
Lennard-Jones and electrostatic potentials between the edge atoms.
  
The force field parameters were fitted to the 5565 ab initio points with a
simplex algorithm \cite{Nelder:1965}. The ab initio energies for the
unbound region were reproduced with a root-mean-square-deviation
(RMSD) of $\approx 0.013$~eV. It was considerably more difficult to obtain a good fit for
the bound state due to strong angular dependence of the potential. The
final fit had an RMSD of $\approx 0.06$~eV by using two minimal
symmetrized sets of FFs (twice 8, altogether 16 FFs). Figure \ref{irc} shows an example of the quality of the
fit. The well depth of the bound state was exactly reproduced by the
global surface. In the following, the bound state FFs are numbered
1--16, whereas those for the unbound states are labeled 17--20 and 21-24,
respectively.
 
\subsection{Molecular-dynamics (MD) simulations}
For the quasi-classical trajectory calculations a code with provisions
for adiabatic reactive MD (ARMD) \cite{Danielsson:2008} was used. The
Hamiltonian equations of motion were solved in Cartesian coordinates
using the adaptive timestep Modified Extended Backward Differentiation
Formulas method (MEBDFSO) \cite{abdulla01}.\\

\noindent 
The dynamics was initiated in the unbound state (FFs 21-24). Between
centre-of-mass separations of $R = 7.56~a_0$ and $R = 7.09~a_0$
the momentarily active unbound surface was smoothly switched to the
bound FFs (numbers 1-16). The complex is considered to be formed for
$R \leq 7.09~a_0$ and the dynamics is continued in the bound
state. When a separation of $R=7.56~a_0$ is reached again, it is
determined which unbound FF is lowest in energy at the corresponding
geometry and the dynamics is followed on this FF by smoothly switching
from the bound state. Following the dynamics further, the system
either can return to the bound state (recrossing) or it can decay into either
N$_2$+N$_2^+$ (charge preserving, FFs 21 to 24) or N$_2^+$+N$_2$
(charge transferring, FFs 17-20). The dynamics is followed until
either the lifetime of the bound state exceeded 100 ps or the initial
separation of the two fragments was reached again. Once the complex
survives a few vibrational periods, energy is expected to have
randomized completely. Our results
show that if the complex survives for at least 0.4 ps the correlation between
the reactant and product states is negligible.

\subsection{Analysis of final sates}
As a consequence of using classical dynamics, products are formed with
rovibrational energies and angular momenta which correspond to
fractional quantum numbers. Also, products having less than zero-point
vibrational energy (ZPE) can be formed. This ZPE leakage is a
shortcoming of quasi-classical simulations and various methods were
proposed for either avoiding or correcting it
\cite{bowman89zpl,miller89zpl,hase94zpl}. In the present study, only
trajectories which satisfy certain constraints in the product states, i.e., the total ZPE of the two molecules should be conserved within $\pm 10\%$, were further analyzed.


\section{Results and Discussion}
\label{results}

\subsection{Product rotational-state distributions}

\begin{figure}[t]
\begin{center}
\epsfig{file=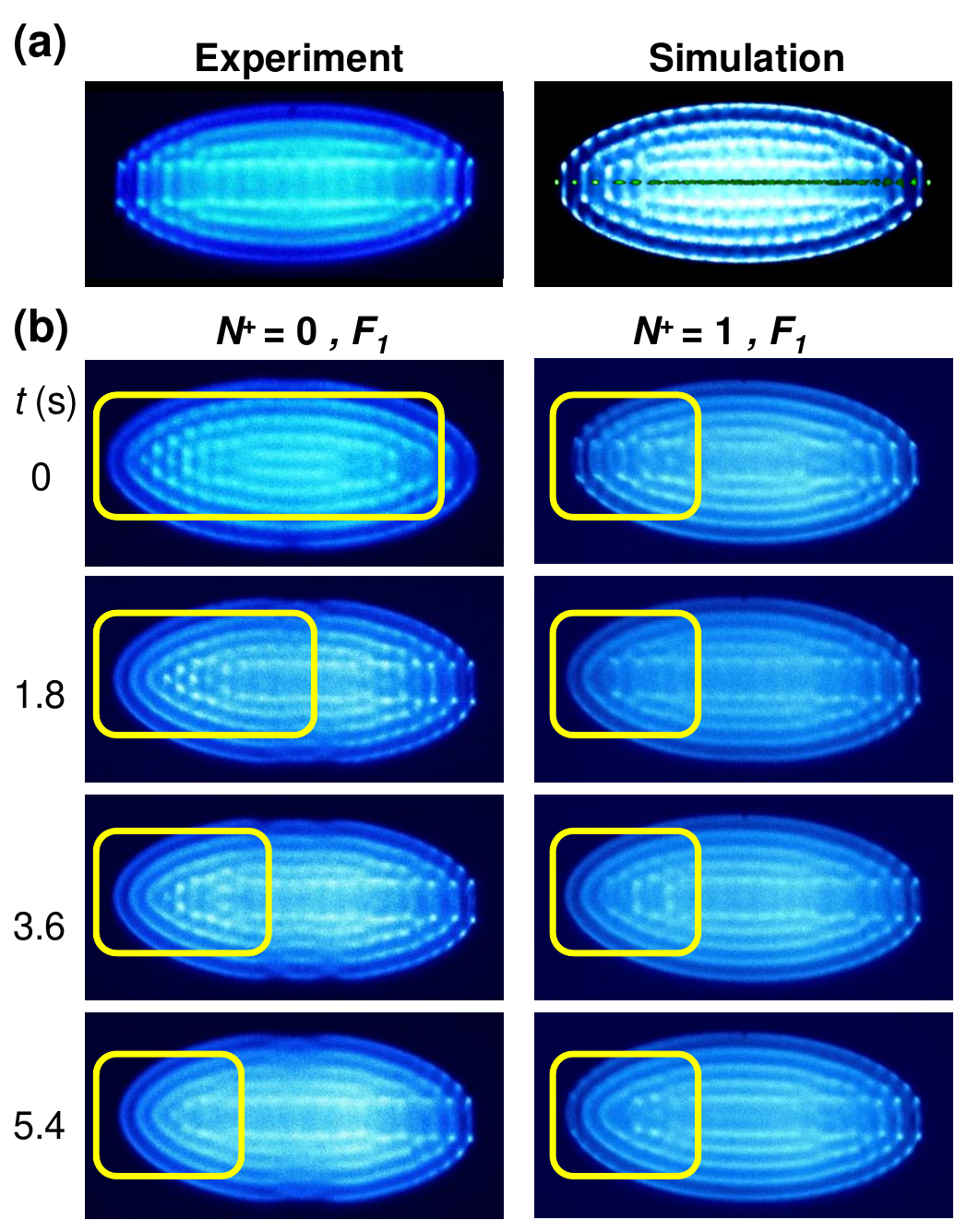,width=0.9\columnwidth}
\end{center}
\caption{(a) Experimental false-color fluorescence image and molecular-dynamics
  simulation of a Ca$^+$/N$_2^+$ bicomponent Coulomb crystal before reaction. The
  spatial distribution of the non-fluorescing molecular ions has been
  made visible in green in the simulated image. (b) LICT experiments probing the populations in the
  N$_2^+~N^+=0, F_1$ and $N^+=1, F_1$ spin-rotational states. The
  panels show fluorescence images obtained after LICT as a function of
  the effective time of reaction $t$ with neutral N$_2$ molecules from
  the molecular beam. The LICT efficiency (corresponding to the number
  of non-fluorescing ions removed from the crystal in the boxed areas)
  decreases in $N^+=0$ and increases in $N^+=1$ with increasing
  reaction time, indicating the generation of rotationally excited
  N$_2^+$ ions from the CT reaction N$_2^+$+N$_2$. See
  text for details.}
\label{LICT}
\end{figure}

Fig. \ref{LICT} (b) shows LICT measurements probing the populations in
the $N^+=0, F_1$ and $N^+=1, F_1$ spin-rotational states as a function
of the effective time of reaction with neutral N$_2$ molecules from
the molecular beam. For the $N^+=0$ state, a marked decrease of the
LICT efficiency can be observed with increasing reaction time,
indicating the removal of the initially prepared N$_2^+,~N^+=0$ ions
by CT with N$_2$. Conversely, an increase in the LICT
efficiency out of the $N^+=1$ state can be observed, indicating the
generation of ions in rotationally excited states as a consequence of
the reactive collisions.

\begin{figure}[t]
\begin{center}
\epsfig{file=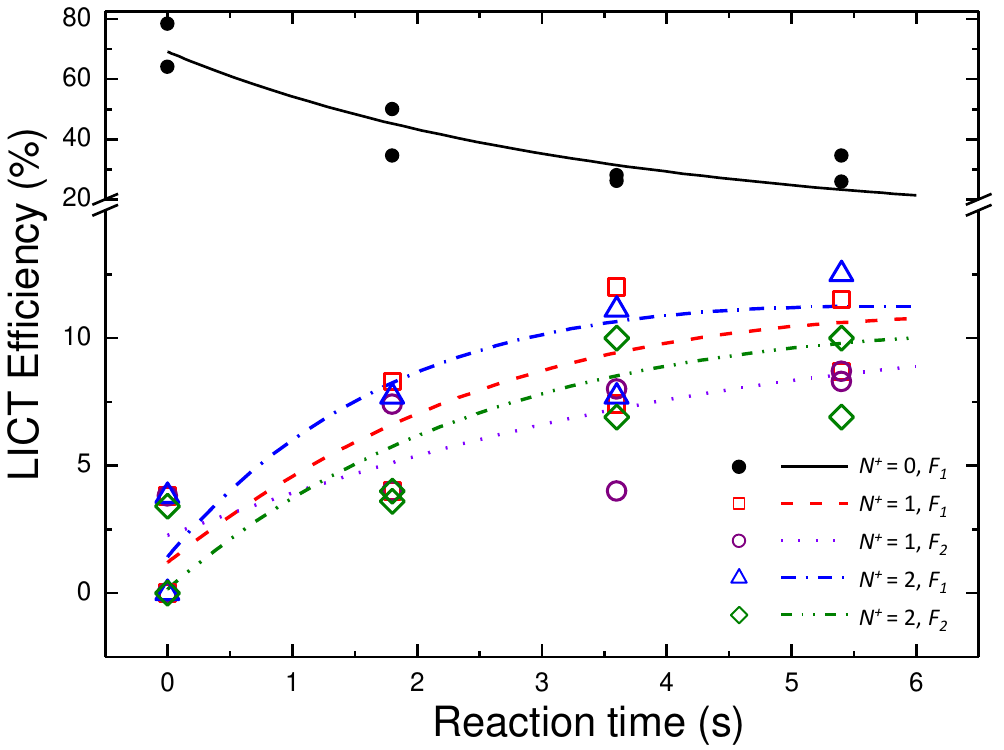,width=\columnwidth}
\end{center}
\caption{\label{pops} LICT efficiencies reflecting the populations in
  the lowest five spin-rotational levels of N$_2^+$ as a function of
  the reaction time with neutral N$_2$ molecules. Over the course of
  the reaction, the initial ensemble of state-selected N$_2^+$ ions is
  replaced by rotationally excited product ions. The lines are a fit
  of the data to the kinetic model shown in Fig. \ref{models} (a).}
\end{figure}

LICT measurements on the population of the $N^+=0,1,2 \ F_{1,2}$
levels as a function of the reaction time are shown in
Fig. \ref{pops}. Two independent LICT measurements were performed for
each spin-rotational level at four different effective reaction
times. As a general trend, a decrease of the population in $N^+=0$ in
favor of an increase of the population in $N^+\geq1$ was
observed. This result indicates that product ions over the whole range
of rotationally excited states probed in the present study were
generated as a consequence of CT between N$_2^+$+N$_2$. Because the neutral
products leave the trap, no conclusions on their
rotational-state distribution can be drawn.

The observed redistribution of population in the ensemble of
Coulomb-crystallized N$_2^+$ ions is the result of a sequence of
CT reactions with neutral molecules from the beam. The
initially state-selected N$_2^+$ ions undergo CT
collisions with rotationally cold N$_2$ molecules. The resulting
N$_2^+$ product ions are sympathetically cooled into the Coulomb
crystal whereupon they in turn can undergo collisions. In this way, the ensemble of originally state-selected
N$_2^+$ ions in the Coulomb crystal is lost and replaced with
rotationally excited product ions.

Close inspection of the data in Fig. \ref{pops} reveals two important
details. First, already in the beginning of the experiment when most
of the reactive collisions occur with ions in $N^+=0$, the generation
of ions in all rotational levels probed in the present study can be
observed. This observation suggests that the production of product
ions is feasible over a broad range of rotational states by reactions
with N$_2^+$ in $N^+=0$. The relevant rates
appear to be of a similar magnitude, at least over the range of
product states probed in the present study.  Second, for a specific
rotational level $N^+$, there appears to be a slight preference for
the generation of the $F_1$ spin-rotational component in comparison to
the $F_2$ component.

\begin{figure}[t]
\begin{center}
\epsfig{file=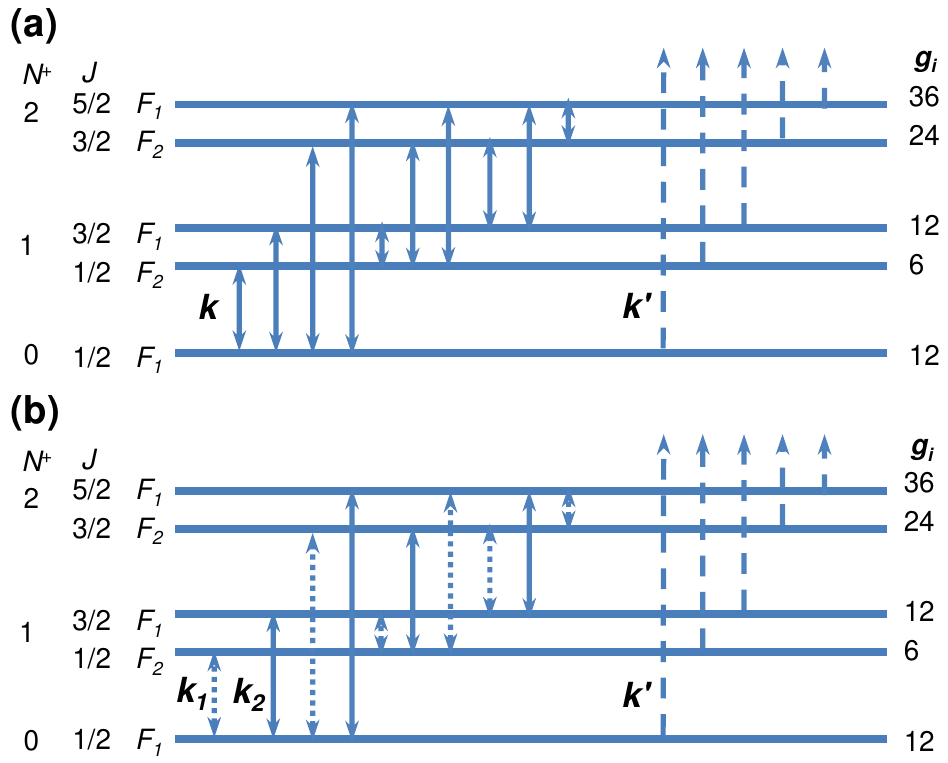,width=0.9\columnwidth}
\end{center}
\caption{\label{models} Schematic of kinetic models used to fit the
  redistribution of rotational populations in the ensemble of
  sympathetically-cooled N$_2^+$ ions as a consequence of CT
  collisions. Model assuming (a) an uniform rate
  constant $k$ and (b) different rate constants $k_1$ and
  $k_2$ for processes connecting unlike and like spin-rotational
  levels $F_{1,2}$ in the reactant and product ions, respectively. See
  text for details. $g_i$ denotes the degeneracies of the
  spin-rotational levels including nuclear-spin statistical weights.}
\end{figure}

\subsection{Kinetics}
\label{kinetics}

The data were analyzed in terms of kinetic models taking into account
CT collisions of the initially prepared N$_2^+ ~(N^+=0)$
reactant ions as well as consecutive reactions of the product N$_2^+$
ions, see Fig. \ref{models}. Because the number density of neutral
N$_2$ in the molecular beam is much larger than the number density of the
N$_2^+$ ions in the trap, pseudo-first-order kinetics was assumed in
both models.

Based on the observation that all product rotational states probed in
the present study ($N^+\leq 2$) seem to be produced with similar
rates, it was assumed in a first model that all processes of the form N$_2^+(N^+)+$N$_2(J)
\rightarrow~$N$_2(\bar{J})+$N$_2^+(\bar{N}^+)$ (bars indicate the
rotational quantum numbers after the decay of the reaction complex) occur with the same
 $k$, see Fig. \ref{models} (a). The
complete loss of population from the manifold of states with $N^+\leq
2$ was taken into account by an effective loss rate constant
$k^\prime$. In the absence of detailed information on the neutral product state distribution, $k$ and $k^\prime$ represent effective rate constants averaged over all neutral product states. The spin-rotational level populations $n_i$ were obtained from
the set of rate equations
\begin{equation}
\frac{\mathrm{d} n_i}{\mathrm{d} t}=k\sum_{j\neq i}(g_in_j-g_jn_i)-k^\prime n_i,
\end{equation}
where $g_{i,j}$ represent degeneracy factors (see Fig. \ref{models}). The
initial state populations $n_i(t=0)$ were also treated as a fit
parameter, accounting for the uncertainty in determining the starting
time of the measurement following the re-alignment of the molecular
beam after ion loading (see Sec. \ref{expt}). The fit of this model to
the experimental data yielded the pseudo-first-order rate constants
$k=0.0023(4)$~s$^{-1}$ and $k^\prime=0.04(3) $~s$^{-1}$. Because the
density of neutral molecules in the molecular beam is not precisely
known, it was not possible to obtain second-order rate constants from
these results. Previous rate measurements with thermal samples of
N$_2$ gas leaked into the chamber \cite{tong11a} were consistent with
the result obtained by Frost et al. \cite{frost94a} that the total
CT rate amounts to one half of the Langevin collision
rate $k_L=8.3\times10^{-10}$~cm$^3$~s$^{-1}$.

The time-dependent populations of the $N^+=0,1,2, F_{1,2}$ states
computed with the kinetic model using the fitted parameters are shown
in Fig. \ref{pops}. The agreement between calculated and
experimental level populations is satisfactory given the spread of the
measured level populations. The agreement vindicates the assumption
that the state-specific rate constants do not (or only weakly) depend on the rotational
state of the product ion within the uncertainty limits of the present
measurement and the range of states studied. The kinetic model also
reproduces the preferential generation of ions in the $F_1$ compared
to the $F_2$ spin-rotational components of the same $N^+$ state
suggesting that this effect is caused by the higher statistical
weight of the $F_1$ components.

To check the validity of this model, its main simplification, i.e.,
the restriction to a ''universal'' state-to-state rate constant $k$,
was relaxed. In a second kinetic model (see Fig. \ref{models} (b)),
different rate constants $k_1$ and $k_2$ for reactions connecting
states with different and like spin-rotational labels $F_{1,2}$,
respectively, were assumed to account for the preferential production
of product ions in the $F_1$ levels. The fit yielded rate coefficients
$k_1=0.0023(9)$~s$^{-1}, k_2=0.0022(7)$~s$^{-1}$ and
$k^\prime=0.04(3)$~s$^{-1}$. The results obtained for $k_1$ and $k_2$
agree with each other and with the value of $k$ within the uncertainty
limits, supporting the conclusion that the increased production rates
of ions in the $F_1$ levels is indeed solely caused by the higher
statistical weights associated with these channels.

\subsection{Trajectory calculations}

Quasi-classical trajectory calculations were performed to interpret the experimental findings and gain further
insight into the energy-transfer mechanisms involved in the
reaction. Initial states for the trajectory calculations correspond to
semi-classically quantized states of rotating Morse
oscillators \cite{miller75}. In the experiment the molecules are
exclusively in their ground vibrational state ($v =0$). The initial
rotational angular momentum $N^+$ of N$_2^+$ ions was set to zero as
prepared in the experiment whereas that of N$_2$ was randomly sampled
according to the experimentally measured distribution (see
Sect. \ref{expt}). Initial coordinates and momenta were generated by
randomly sampling the phase-space distribution of the rotating
Morse-oscillators and the spatial orientation of each molecule was
taken from a uniform distribution within $4 \pi$ steradian. The
magnitude of the angular velocity for a given vibrational phase was
set to the angular momentum calculated from $L=(J(J+1))^{1/2}\hbar$
and its direction was randomly sampled from a uniform distribution in
the plane perpendicular to the axis of the molecule.

\begin{figure}[t]
\begin{center}   
\epsfig{file=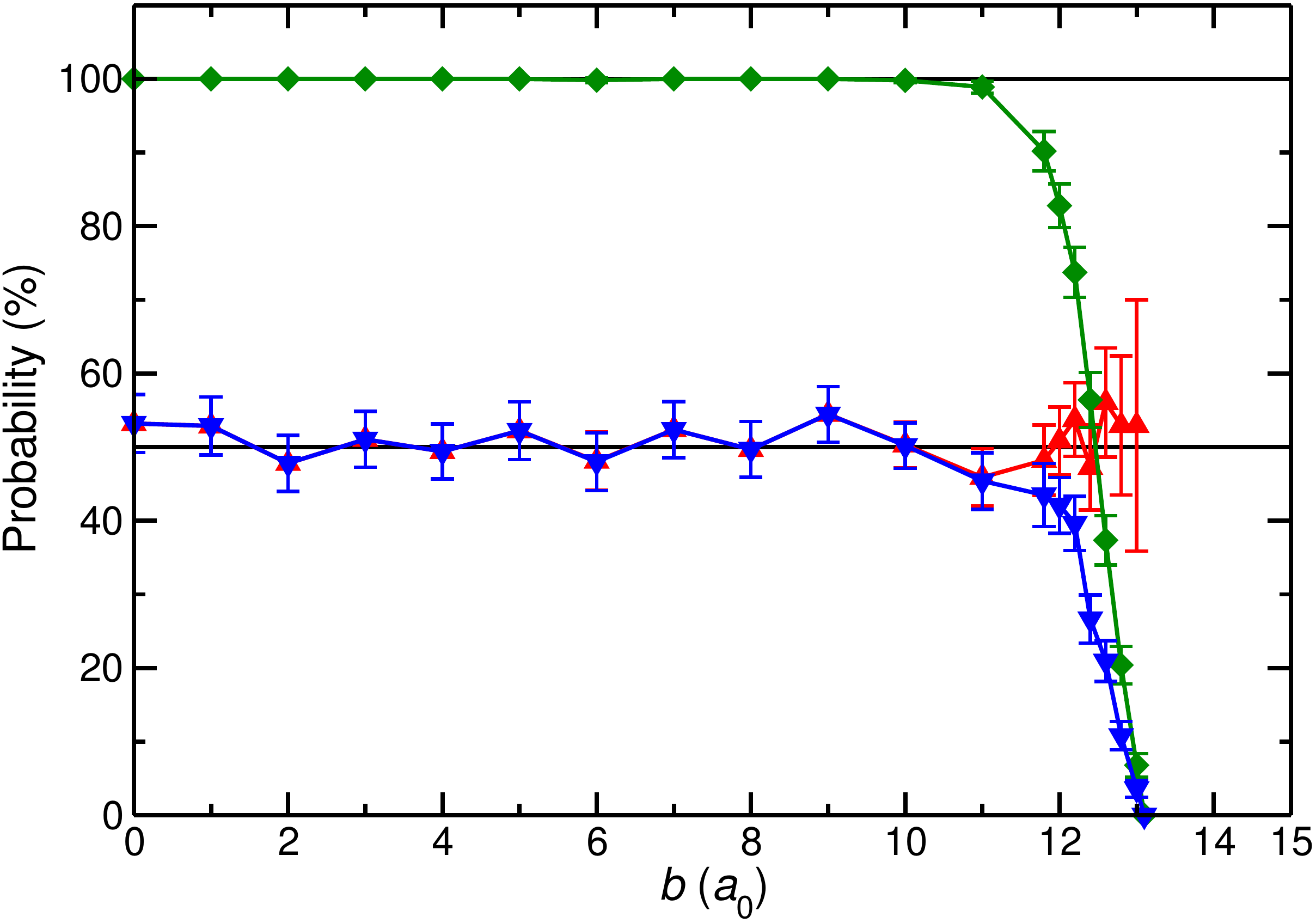,width=0.9\columnwidth}
\end{center}                                
\caption{Opacity functions for complex formation (green diamonds) and
  charge transfer (blue triangles), and probability curve for
  charge transfer (red triangles) if the complex is formed. Error bars
  correspond to a 2$\sigma$ standard deviation.}
\label{reac}
\end{figure}   

Impact parameters $b$ were between 0 and 15~$a_0$ in steps of 1~$a_0$,
and in steps of 0.2~$a_0$ between 12 and 13~$a_0$ where the opacity
function drops to zero (see below). The initial separation of the
center of mass of the two molecules was 20 \AA\/ ($\approx 38~a_0$),
for which the intermolecular interaction energy was less than $1.4
\times 10^{-4}$~eV. According to the Langevin-Gioumousis-Stevenson
(LGS) model of ion-molecule reactions the rate does not depend on the
relative velocity of the partners \cite{Gioumousis:1958}. Therefore,
the relative velocity of the colliding molecules was set to the calculated beam
flow rate (787 m/s) without dispersion.

To validate the simulations, the probability of complex formation and
the probability of CT were determined for a range of
impact parameters $b$ based on 500--1000 trajectories at each value
(see Figure \ref{reac}). According to our quasi-classical model, the
probability for CT was found to be 50\% (within the
statistical uncertainty) up to impact parameters $b=9~a_0$. The CT probability
starts dropping slowly between 10 and 12 $a_0$, and above 12~$a_0$ it
decays steeply to zero around 13~$a_0$. The probability of complex
formation is 100\% up to 9~$a_0$ and drops proportionally to
the CT probability. Once the complex is
formed, the probability for CT is $\approx 50$\%
regardless of the impact parameter. This finding is expected based on symmetry arguments and
considering that rapid charge-redistribution takes place before the
complex decays. It is also in agreement with previous experimental results \cite{frost94a}.

From the integrated opacity function an integral cross section for
CT of $\sigma_{\rm tot} = (243 \pm 19)~a_0^2$ was
obtained. Multiplication with the relative velocity (787 m~s$^{-1}$) yields
$k = (5.36 \pm 0.42) \times 10^{-10}$ cm$^3$s$^{-1}$ for the second
order rate coefficient for CT. This value compares
favorably with previous experimental results of $4.24 \times 10^{-10}$
cm$^3$~s$^{-1}$, $6.6 \times 10^{-10}$ cm$^3$~s$^{-1}$ and $5.0 \times
10^{-10}$ cm$^3$~s$^{-1}$ obtained for the CT reaction between
$^{15}$N$_2^+ (v=0)$ and $^{14}$N$_2(v=0)$ \cite{frost94a,adams81,mcmahon76}.

\begin{figure}[t]
\begin{center}   
\epsfig{file=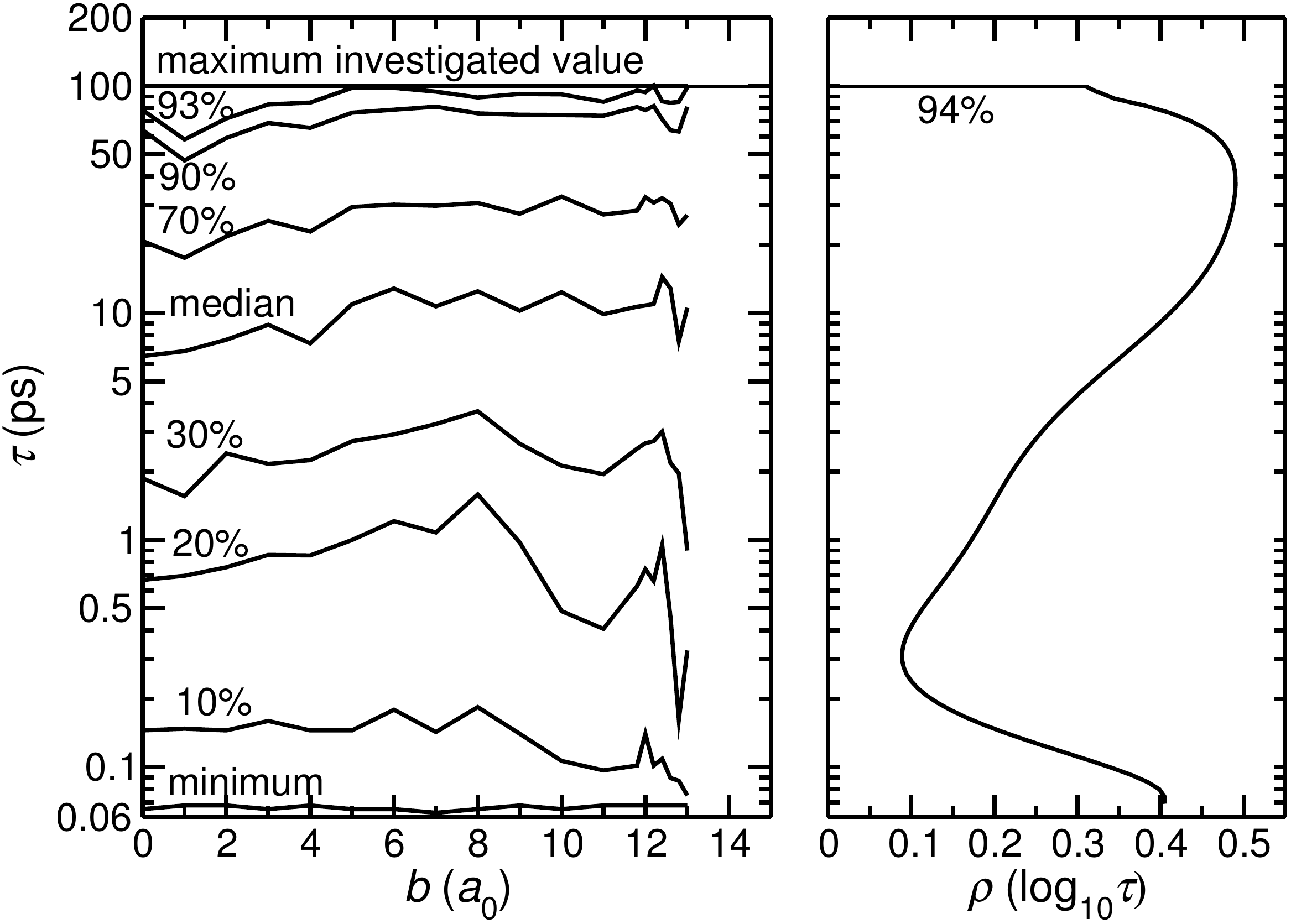,width=0.9\columnwidth}
\end{center}                                
\caption{\label{life} Percentiles representing the distribution of lifetimes $\tau$ for the N$_4^+$ complex as a function of the impact parameter $b$ (left panel) and integrated probability density $\rho$ (right panel). The median lifetime is around 10 ps, the complex decays within 100 ps in $\approx 94$\% of all observed
  cases.}
\end{figure}

The lifetime of the complex was determined as the time difference
between the last and the first crossing of the surface separation
radius $7.09~a_0$ (see Sect. \ref{theo_pes}).  As Figure \ref{life}
suggests, the median lifetime of N$_4^+$ is $\approx 10$ ps and does
not or only slightly depends on $b$. Lifetimes up to 100 ps are found for 94\% of the
trajectories which qualitatively agree with previous
estimates \cite{vankoppen84a, phillips90}.

\subsection{Translation-to-rotation energy transfer}

\begin{figure}[t]
\begin{center}   
\epsfig{file=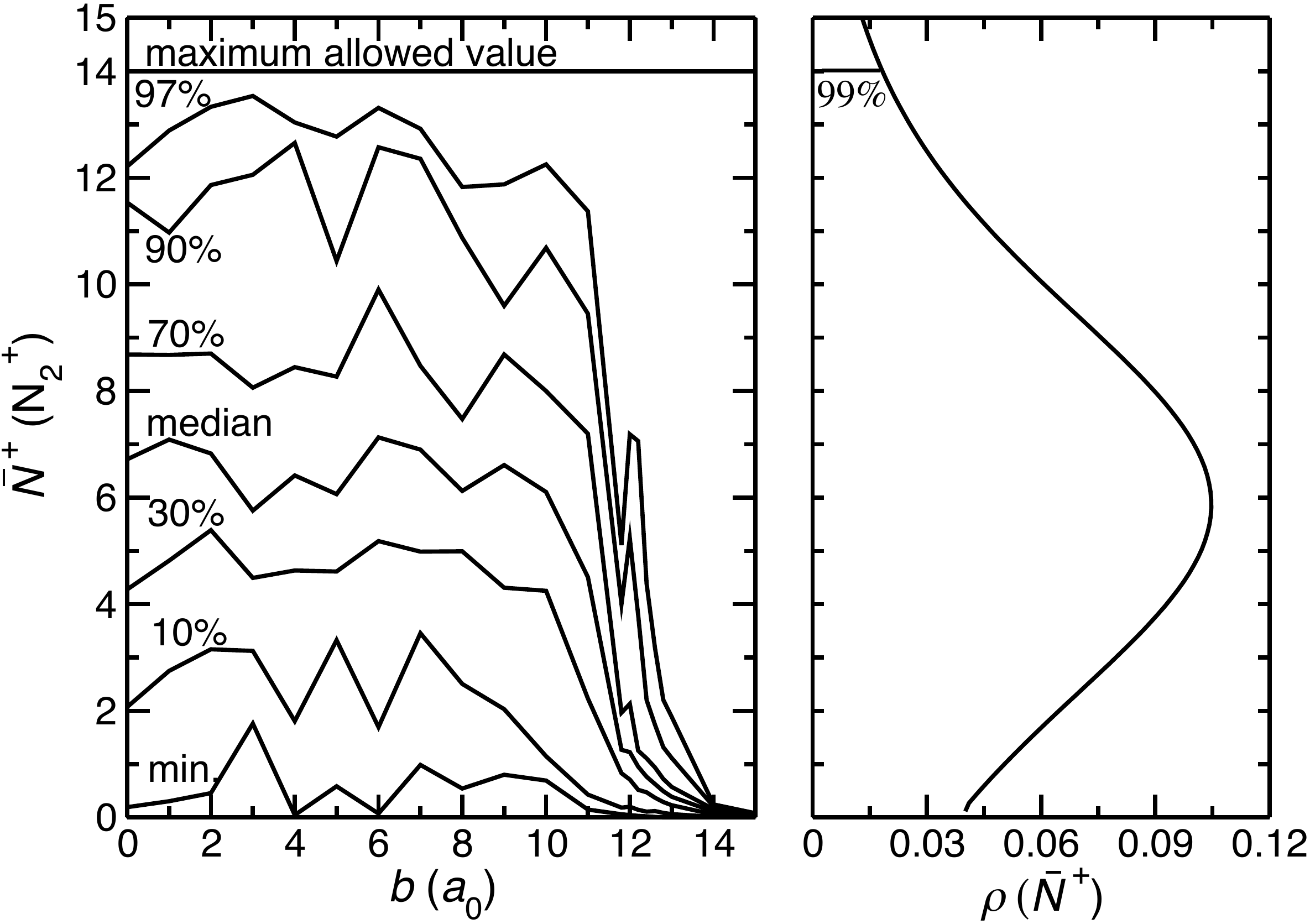,width=0.9\columnwidth}
\end{center}                                
\caption{\label{jdist} Percentiles representing the rotational-state distributions for the N$_2^+$ products after breakup of the reaction complex as a function of the impact parameter $b$ (left panel) and integrated probability density $\rho(\bar{N}^+)$ (right panel). The 
  products are predominantly formed in rotationally excited
  states.}
\end{figure}

The salient quantity which can be extracted from the trajectories to aid in the interpretation of the experimental results is the distribution of rotational angular momentum $\rho(\bar{N}^+)$ of the N$_2^+$ products after complex
formation and decay. Figure \ref{jdist} reports $\rho(\bar{N}^+)$ and
suggests that for the majority (97\%) of cases $\bar{N}^+ \leq 13$. On
energetic grounds, a total of 0.045~eV is available in collisional
energy for the reactants, and the initial rotational quantum number of
N$_2$ can be as high as 3 (corresponding to a rotational energy of $\approx 0.003$~eV), which implies
$\bar{N}^+ \leq 14$. The N$_2^+$ products are predominantly formed in
excited rotational states, which supports the experimental findings of
depletion of the N$_2^+$ ground state population caused by reactive
collisions with the molecules in the beam. Note that the theoretical product state distribution cannot be compared directly with the experimental findings reported in Fig. \ref{pops}, because the N$_2^+$ ions probed in the experiment result from a sequence of CT reactions, whereas the simulations only reflect single collision events.

\begin{figure}[t]
\begin{center}   
\epsfig{file=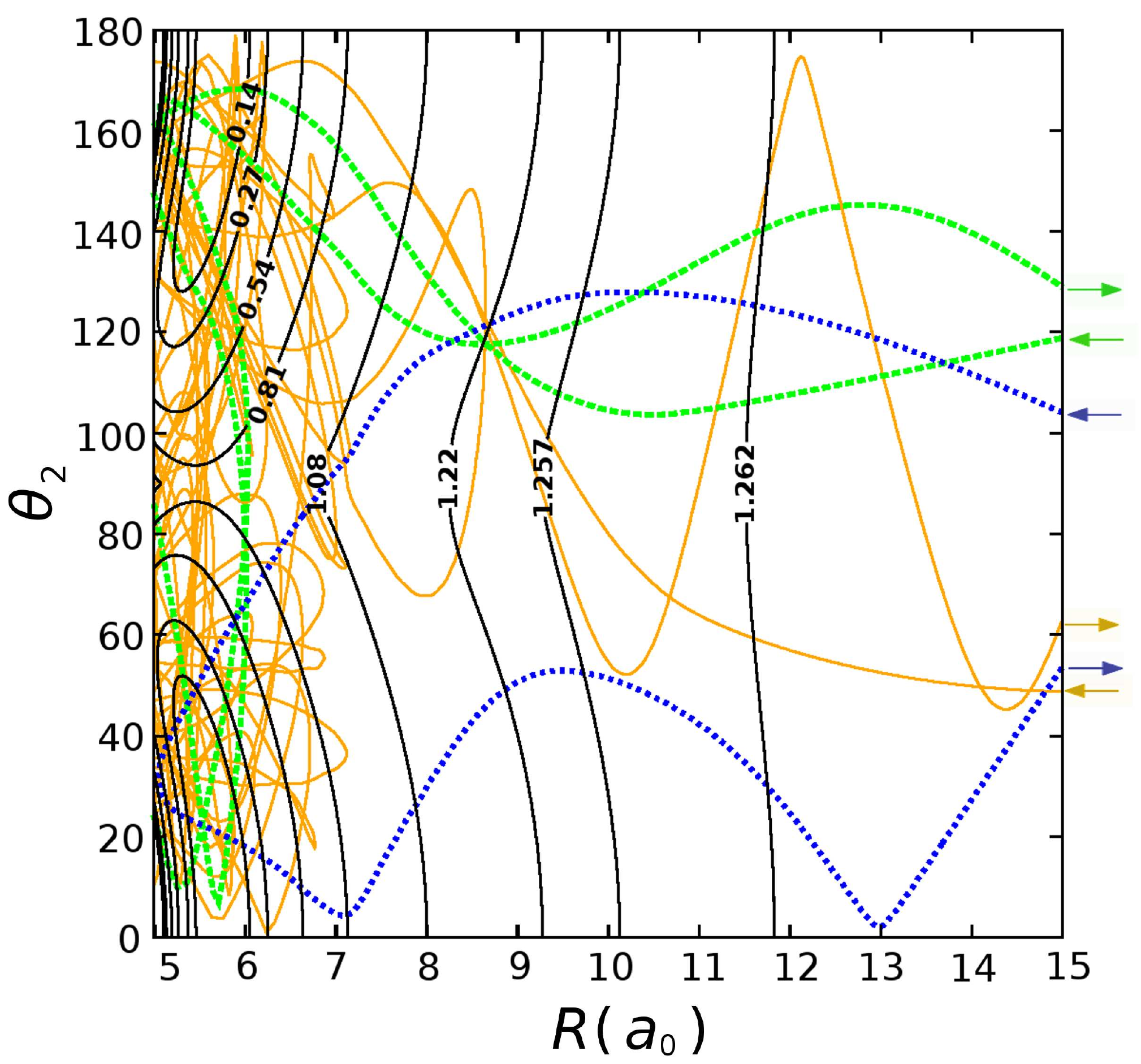,width=\columnwidth}
\end{center}
\caption{Projection of the PES onto the $R$ and $\theta_2$ coordinate
  (see Figure \ref{irc}) with $r_1=r_2=2.08~a_0,\theta_1=0$, and
  $\varphi=0$. Contour lines are drawn between 0 and 1.262~eV. Three
  typical trajectories leading to rotational excitation of the products are reported. The arrows indicate the
  direction of motion (incoming and outgoing).  The complex lifetimes amount to
  87 fs (blue), 341 fs (green) and 6604 fs (orange).}
  \label{pes}
\end{figure}    

However, the trajectory calculations clearly reproduce the rotational excitation of the products observed in the experiment and reveal the mechanisms of the underlying T-R energy transfer occurring during CT. The reactant molecules approach each other in a random orientation and are
accelerated towards the potential well of the linear reaction
complex. The torque exerted while forcing the reactants towards a
linear configuration excites a counter-rotation of the N$_2$ moieties
which results in highly excited bending and torsional vibrations of
the complex. Moreover, anharmonic couplings lead to a practically
complete redistribution of the available energy over all vibrational
degrees of freedom during the lifetime of the complex. Upon its
breakup, large-amplitude bending and torsional vibrations are converted into
product rotations. Representative trajectories, projected on the PES,
are shown in Fig. \ref{pes} to illustrate this effect.

Close inspection of the trajectories reveals the likely presence of
additional processes resulting in the rotational excitation of the
products, including rotation-vibration coupling, back scattering and recrossing
which all may contribute to the final state distribution. Under the
present experimental conditions, the orbital angular momentum of complex-forming collisions is computed to be $L\lesssim 140\hbar$ which is available for conversion
into rotational motion of the reaction complex. Coriolis forces may lead
to a coupling of the complex rotation to its internal motion,
providing another mechanism for the transfer of angular momentum to
the fragments. Furthermore, the trajectories shown in Fig. \ref{pes}
suggest that a wide range of scenarios is possible. They include
direct mechanisms, illustrated by the blue trajectory, and can range
to bound states with almost full redistribution (randomization,
i.e. IVR) of the internal energy as shown for the orange trace. This
is reminiscent of the situation recently encountered in the
vibrationally induced photodissociation of sulfuric acid where all
regimes from prompt reaction to complete IVR were found, depending on
the amount of internal energy made available to the
molecule \cite{MM11h2so4}.


\section{Summary and conclusions}
\label{summary}
In the present study we have demonstrated for the first time
ion-molecule reaction studies with state-selected Coulomb-crystallized
molecular ions using the symmetric CT reaction N$_2^+$+N$_2$ as an
example. By rotational state selection of the reactant ions and
their localization in space by sympathetic cooling in an ion trap,
 their internal and translational motions were completely
controlled. By simultaneously cooling the neutral co-reactants to the
lowest rotational states and collimating their kinetic-energy
distribution in a supersonic molecular beam, ion-molecule reactions were performed with an unprecedented degree of
control over both the collision energy and internal states of the reaction
partners. The analysis of the product-ion rotational state
distribution yielded for the first time information on T-R energy
transfer occurring during CT. The experimental results were analyzed
and interpreted by making contact with quasi-classical trajectory
simulations on a full-dimensional PES. The simulations reproduced the
experimental findings and yielded important insights at an atomistic
level into the mechanisms underlying T-R energy transfer.


\end{document}